\documentclass[useAMS,usenatbib]{mn2e}
\usepackage{graphicx}
\usepackage{subfigure}
\usepackage{amsmath}
\usepackage{amssymb}
\usepackage{color}
\usepackage{natbib}
\usepackage{slantsc}

\title[Spatially resolved star formation histories of nearby galaxies: evidence for episodic star formation
in discs]{Spatially resolved star formation histories of nearby galaxies: evidence for episodic star formation
in discs}
\author[M.-L. Huang et al.]{Mei-Ling Huang$^{1}$\thanks{E-mail:
mlhuang@mpa-garching.mpg.de}, Guinevere Kauffmann$^{1}$, Yan-Mei Chen$^{2,3}$, 
Sean M. Moran$^{4}$, 
\newauthor
Timothy M. Heckman$^{4}$, Romeel Dav\'{e}$^{5}$, Jonas Johansson$^{1}$\\
$^{1}$Max-Planck Institute for Astrophysics, Karl-Schwarzschild-Str. 1, D-85748 Garching, Germany\\
$^{2}$Department of Astronomy , Nanjing University, Nanjing 210093, China\\
$^{3}$Key Laboratory of Modern Astronomy and Astrophysics (Nanjing University), Ministry of Education, Nanjing 210093, China\\
$^{4}$Department of Physics and Astronomy, The Johns Hopkins University, 3400 N. Charles Street, Baltimore, MD21218, USA\\
$^{5}$Astronomy Department, University of Arizona, Tucson, AZ 85721, USA\\
}

\begin{document}
\date{ in original form 2012 September}
\pagerange{\pageref{firstpage}--\pageref{lastpage}} \pubyear{2002}
\maketitle
\label{firstpage}

\begin{abstract}
Using  long-slit spectroscopy from \citet{mor} we constrain the radial 
dependence of the recent star formation histories of nearby galaxies with 
stellar masses greater than $10^{10} M_{\odot}$.  
By fitting stellar population models to the combination of SSFR, D4000 and 
H$\gamma_{A}$, we show that the star formation histories of many disk 
galaxies cannot be accurately represented if their star formation rates 
declined exponentially with time. Many galaxies have 
Balmer absorption line equivalent widths that require recent short-lived 
episodes or bursts of star formation. 

The fraction of galaxies that have experienced episodic rather than continuous 
star formation is highest for ``late-type'' galaxies  with low stellar masses, 
low surface densities, and low concentrations. In these systems, bursts occur both
in the inner and in the outer regions of the galaxy.  
The fraction of stars formed in a single burst   
episode is typically around 15\% of the total stellar mass in the 
inner regions of the galaxy and around 5\% of the mass in the outer regions. 
When we average over the population, we find that such bursts contribute 
around a half of the total mass in stars formed in the last 2 Gyr. 
In massive galaxies, bursts occur predominantly in the outer disk.
Around a third of all massive,  bulge-dominated galaxies 
have experienced recent star formation episodes that are fully 
confined to their outer ($R> 0.7 R_{90}$) regions. 
The fraction of stars formed in bursts is 
only $\sim 2-3$ \% of the underlying stellar mass, but when we 
average over the population, we find that such 
bursts contribute nearly all the stellar mass formed in the last 2 Gyr.  

Recent star formation in outer disks is strongly correlated with the
global atomic gas fraction of the galaxy, but not its global molecular gas fraction.
We suggest that outer episodic star formation is triggered by
gas accretion ``events''. Episodic star formation in the inner regions  is suppressed  
in galaxies with large  bulge-to-disk ratio. This supports the idea that inner bursts are
linked to instability-driven gas inflows.  

\end{abstract}


\section{Introduction} 

It is now well-established that galaxies in the nearby Universe separate
rather cleanly into two classes: those with disky morphologies, plentiful
gas and ongoing star formation and those that are bulge-dominated, with
little gas and star formation, and where star formation has largely ceased
(e.g. \citealt{str};  \citealt{kaub}; \citealt{bal}).
Most of the baryons in the Universe are not locked up in stars, but reside
in the circum-galactic medium associated with virialized dark matter halos
or in a more diffuse phase in the intergalactic medium \citep{fuk}. It is thus
reasonable to postulate that the observed bimodality in galaxy properties
is in some way linked with the ability of galaxies to accrete gas from the
external environment.

It is very difficult to probe gas infall onto galaxies directly. Resolved
studies of the stellar populations of galaxies may, however, provide some
important clues as to how accretion may refuel galaxies. \citet{bela} 
used a large sample of low inclination spiral galaxies with radially
resolved B,R,K photometry to derive estimates of age and
metallicity as a function of position within the galaxy. \citet{belb}
then interpreted these results using very simple models of disk
galaxy evolution. They showed that a ``closed box'' model in which the star
formation history and the metallicity of a given area in any galaxy depend
only on the initial local gas surface density did not provide a good fit to
the observational results. Real galaxies had stronger age gradients than
predicted, and the predicted slope of the age/central surface brightness correlation
was steeper than observed.  Models which included infall of gas (with an
infall rate that was larger for low mass galaxies) were required
to fit the age/surface brightness relations.  Outflows were required to
explain the observed correlation between
stellar metallicity and galaxy mass.

\citet{mun} examined the specific star formation rate profiles
of a sample of nearby, face-on spiral galaxies with UV photometry from the 
GALEX Atlas of Nearby Galaxies \citep{gil} and K-band photometry from the 
Two Micron All Sky Survey \citep{skr}. They showed that on average,
galaxies were forming stars at higher relative rates in their outer regions compared
to their inner regions.  
More recently, \citet{wan} showed that UV/optical colour gradients
of galaxies are strongly correlated with their atomic gas content. 
Gas-rich galaxies with high H\textsc{i} gas masses had stronger colour gradients than
galaxies of the same stellar mass, size, and
NUV-r colour with average H\textsc{i} content. Both \citet{mun}
and  \citet{wan} interpreted their results in the context of an
``inside-out'' picture of disk galaxy formation, which has commonly served as
a basis for semi-analytic models of the formation of discs in the context
of cold dark matter cosmologies.

The UV/optical colours of galaxies depend strongly on the amount of dust
extinction in the galaxy and also are sensitive to star formation
occurring over timescales of $\sim$1 Gyr \citep{kon}. They thus cannot be used to
assess whether the recent star formation history of a galaxy has been
smooth or episodic. Combinations of narrow-band stellar absorption line indices, in particular
the 4000 \AA\ break strength and the equivalent widths of Balmer absorption lines
, such as H$\gamma$ or H$\delta$, are able to probe the recent star formation
histories of galaxies in more detail. Galaxies that have undergone a burst of
star formation in the last 1-2 Gyr will have stronger Balmer 
absorption line equivalent widths
for a given value of 4000 \AA\ break strength, compared to galaxies that have been
forming their stars continuously (\citealt{dre}; \citealt{pog}; \citealt{kaua}).

This paper aims to constrain the radial dependence of the recent star 
formation histories of disk galaxies. We compare spectral indices, 
the 4000$\AA$ break strength (D$_n$4000), a Balmer line index derived using an approach
based on principal component analysis (PCA), and  
the present-day star formation rate over stellar mass 
(specific star formation rate ; SSFR) to a library of models generated from the \citet{bru} population 
synthesis code. We use the best fit models to constrain the timescale 
over which stars have been formed at different radii in the disk.  
We show that the star formation histories of 
many spiral galaxies cannot be be described by simple exponentially 
declining star formation histories. Additional recent ($<$ 2 Gyr old) 
episodes of star formation are required 
in order to fit D$_n$4000, H$\gamma_A$ and SSFR simultaneously. 
The mean fraction of stars formed in these recent episodes  
is similar in the inner and outer regions of galaxies, except for massive
galaxies, where the bursts occur predominantly in the outskirts
of the galaxy. 

We also study how the recent star formation histories and gas-phase metallicities
in the inner and outer regions of the galaxies in our sample correlate with their
atomic and molecular gas content. Our main result is that the recent star formation
history and the metallicity in both the inner and  outer regions of galaxies 
are strongly correlated with their total atomic gas  mass fraction. In contrast, 
only the inner star formation history and metallicity are correlated
with total molecular gas fraction.  
   
Our paper is organized as follows. Section 2 introduces the 21cm and CO(1-0) line   
data we use from GASS \citep{cat} and COLD GASS \citep{sai} 
surveys and the long-slit spectroscopic observations from the MMT follow-up 
observations \citep{mor}. We explain how we create a library of model 
spectra and use it to extract estimates of parameters such as burst mass fraction in
Section 3. Based on the results of the fits,
we classify our sample galaxies into five groups according whether   
their star formation histories are best described by continuous or
burst models in their inner and outer regions.
In Section 4, we examine how the fraction of galaxies in these different classes
depends on their location in the plane of stellar mass versus 
stellar surface density/concentration/colour. We also examine how the fraction
of stars formed both continuously and in bursts in the inner and outer regions of
the galaxy depends on the location of galaxies in these planes.  
We summarize and discuss our results in Section 5.

\section[]{Data} 

The data set on which this analysis is based has been described in detail in
\citep{mor}. The original parent sample consists of galaxies
with H\textsc{i} line flux data from the Arecibo telescope that were
observed as part of the GALEX Areceibo SDSS (GASS) survey (Catinella et al.
2010; 2012). 
GASS aimed to measure the neutral hydrogen content for a large, uniform sample 
of $\sim$1000 massive galaxies with stellar masses in the range  $10^{10}-10^{11.5} M_{\odot}$ 
and redshifts in the  range $0.025 < z < 0.05$. No other selection criteria such as
cuts on morphology, colour or gas content were applied. This sample thus provided 
an unbiased view of how the cold gas fraction related to the physical parameters 
of galaxies. The reader is referred to these papers for details about the
sample selection and the HI observations.
A subset of 300 galaxies were selected from this sample for CO ($J=1-0$)
observations using the IRAM 30m telescope \citep{sai}. 
Of these, around 200 galaxies now have reduced long-slit spectroscopy
obtained from both the Blue Channel Spectrograph on the 6.5m MMT telescope on Mt.
Hopkins, AZ, and the Dual Imaging Spectrograph on the 3.5m telescope at
Apache Point Observatory (APO; see \citealt{mor10} for details about the
observational setup and data reduction).
These galaxies were all selected at random from the GASS sample.

The wavelength ranges for MMT and APO observations are $\sim$3900-7000 \AA 
\ at a spectral resolution of $\sim$4 \AA\  (90 km s$^{-1})$, and
$\sim$3800-9000 \AA\  at a spectral resolution of $\sim$6-8 \AA\  
(150 km s$^{-1}$).
The spectra were spatially binned outward from the galaxy center to 
ensure an adequate $S/N$ in each spatial bin. All bins have minimum 
extent equal to the slit width: 1.25$\arcsec$ for the MMT observations) and  1.5$\arcsec$  
for the APO observations. The bin size is typically $<$3$\arcsec$ , corresponding to a physical size of 
$\sim$1.5-3 kpc at the redshift of the galaxies in our sample.  The $S/N$ per bin is required  to be 
larger than 15 (per \AA) in the inner regions of the galaxy. In the
outer, lower surface brightness regions, these requirements are progressively loosened 
in order to compromise between  $S/N$ and spatial resolution: 
we adopt  $S/N$ $>$10 for bins out to a distance of 4.5$\arcsec$, $S/N$ $>$8 for bins 
between 4.5 and  6$\arcsec$ and finally $S/N$ $>$6 in the outermost bins.
The spectra were flux-calibrated by matching to SDSS $g$ and $r$-band 
photometry measured through an aperture matched to the slit. 
The O3N2 index \citep{pet} was applied to estimate gas-phase metallicities. 
The empirical relation is 12 + log(O/H) = 8.73 - 0.32 O3N2,
where O3N2 = log(([O\textsc{iii}]/H$\beta$)/([N\textsc{ii}]/H$\alpha$)).
It was demonstrated that the inferred metallicity gradient did
not depend on the exact choice of indicator.

All galaxies have NUV and FUV fluxes from the GALEX medium imaging survey. 
SDSS $r$-band and GALEX NUV images were used to derive the NUV-r colours of
the galaxies in  our sample.
The SDSS images were first degraded to the resolution of NUV images. The $r$-band and 
NUV magnitudes within Kron elliptical apertures were obtained using 
SExtractor \citep{ber} and corrected for Galactic extinction following \citet{wyd}. 
The reader is referred to \citet{cat} for a more  detailed explanation.
Global galaxy parameters such as stellar mass, stellar surface mass 
density, and concentration index were taken from the MPA/JHU value-added 
catalogs (http://www.mpa-garching.mpg.de/SDSS).
Stellar masses were derived by fitting SDSS photometry to stellar population synthesis models.
Stellar surface mass density was defined as $M_{*}/(2\pi R^{2}_{50,z})$, where $R_{50,z}$ was 
the radius containing 50\% of the Petrosian flux in $z$-band. 
Concentration index was defined as $R_{90}/R_{50}$ where
$R_{90}$ and $R_{50}$ are the radii enclosing 90\% and 50\% of the total $r$-band light.

\begin{figure*}
 \includegraphics[width=157mm]{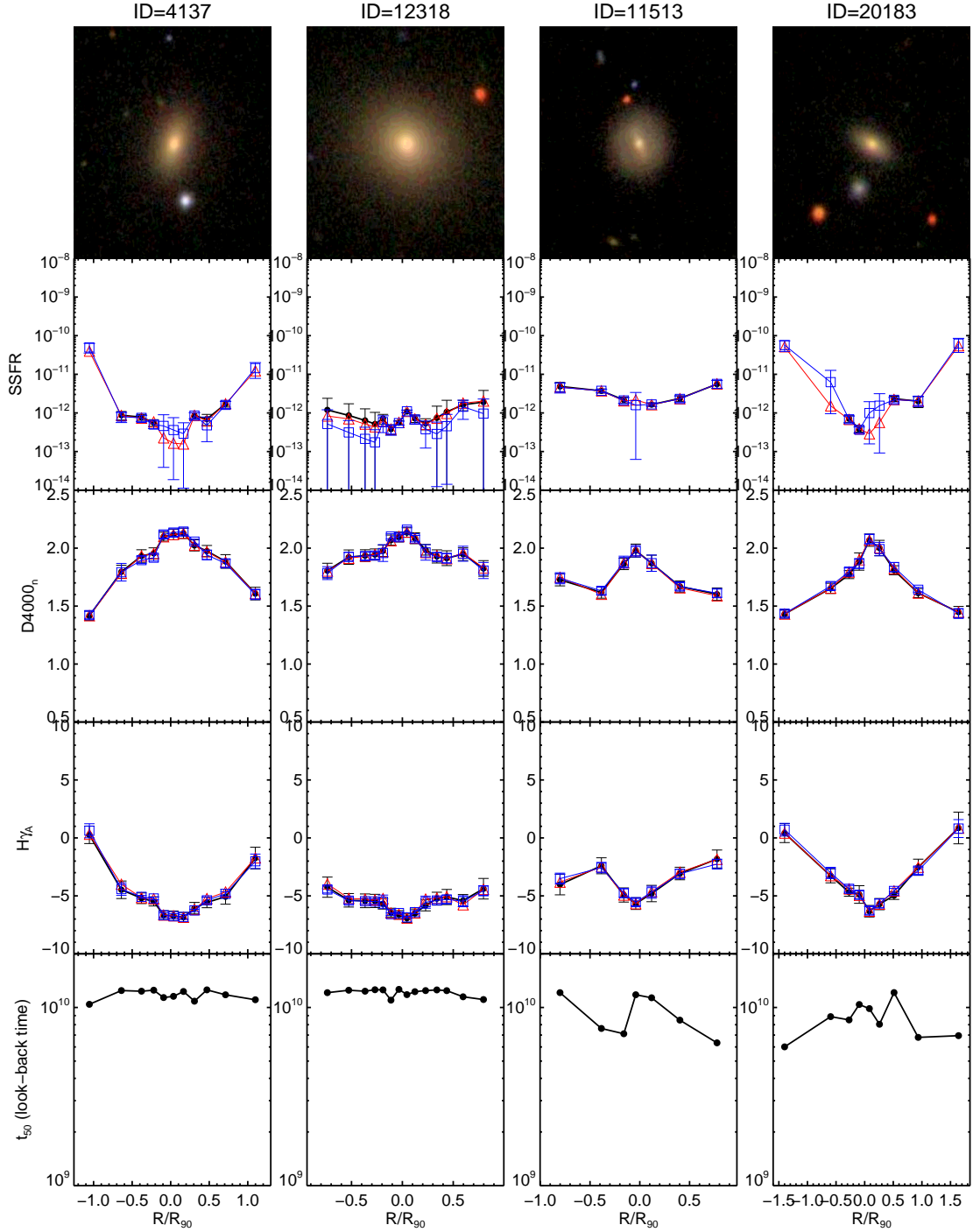}
    \caption{Model fits for galaxies with star formation histories of  Type A
(columns 1 and 2) and Type B (columns 3 and 4). The first row shows SDSS cut-out images
of the galaxy. Rows 2,3, and 4 show the radial run of the specific star formation rate,
the 4000 \AA\ break index and the H$\gamma_A$ index. Black points show the data along with
error bars. Red/blue lines are the model fits using minimum chi-square model and the median of the 
PDF, respectively. In the bottom row we plot the look-back time when half of stellar mass was 
formed. Note that radii
are scaled by dividing by R$_{90}$, the radius enclosing 90\% of the $r$-band light.}
  \label{typeAB}
\end{figure*}

\begin{figure*}
 \includegraphics[width=157mm]{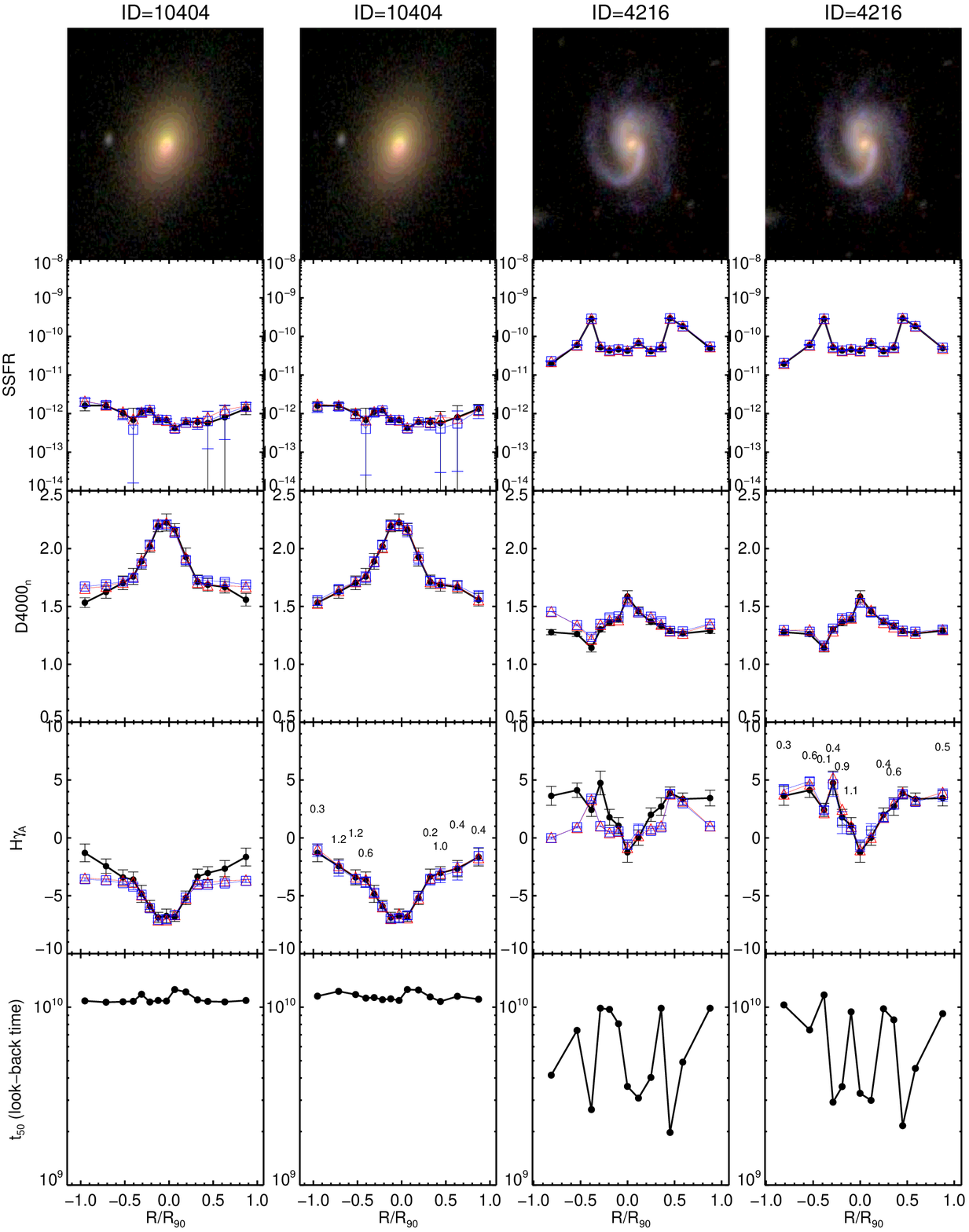}
    \caption{Examples of a galaxy of Type C (Columns 1-2) and Type D ( Columns 3-4).
The rows and line styles have the same meaning as in Figure 1.   
Columns 1 and 3 show the best-fit continuous models. Columns 2 and 4 show the
best-fit models that include bursts. The number above the bins in the fourth row 
is the look-back time of the start of bursts in units of Gyr.}
  \label{typeCD}
\end{figure*}

\section[]{Analyses} 

\subsection{A PCA-based Balmer absorption line index} 

The use of ``classic" Balmer absorption line Lick indices such as H$\delta_A$ or H$\gamma_A$ 
measured directly from the galaxy spectra can be problematic for two reasons: 
1) Because these indices are defined over a narrow range in wavelength, the signal-to-noise
of the measurements is quite low, especially towards the outer regions of the galaxy, where
the spectra may only have  $S/N \sim 2$ per pixel. 
2) In star-forming galaxies, the Balmer absorption lines are filled in by emission.
A robust measurement of the absorption line equivalent width hinges on subtracting
this emission accurately. 

If stellar population models fit the observed spectra accurately, they provide one way
of solving the problem. 
Because the  models are fit to the spectra over a wide range in wavelength,
the errors on Balmer line indices derived from these fits decrease
with respect to those for directly measured indices \citep{wil}.  
In addition, the wavelength regions where emission may be present are masked 
when the fitting is carried out.

In this paper, we apply the principal component analysis (PCA) method
described in \citet{che}, which is an extension of the method
described in \citet{wil}. The 
principle of PCA method is that a galaxy spectrum can be decomposed into a set 
of orthogonal principal components (PCs), which can be linked to physical 
properties such as fraction of stars formed over the last Gyr, or to
estimates of indices such as  D4000, H$\delta_A$, H$\gamma_A$ using a library of
star formation histories generated using population synthesis models.    
By comparing the amplitudes of the principal components  of the observed spectra to those of 
the modelled spectra, the likelihood distribution of such parameters can be 
calculated and thus constrained.  
The advantage of this approach is that it improves our estimates
of Balmer absorption line equivalent widths, particularly for the spectral bins with
poor $S/N$ in the outskirts of the galaxies. The reader is referred to \citet{che} for
further details. In this work, we find the linear combination of 
PC amplitudes that best represents the indices D4000 
and H$\gamma_{A}$ for each spectrum.  A probability distribution function 
(PDF) for either D4000 or H$\gamma_{A}$ can be built using 
$\chi^2$ as weights. We use the median value of the PDF  as our nominal estimate of the 
parameters, and the 16th to 84th percentile range of the PDF as the $\pm 1\sigma$ 
confidence interval.

\subsection{Comparison with a Library of SEDs of Model Galaxies} 

We compare two PCA-derived spectral indices, which we will denote PCA$_{\rm D_n(4000)}$ 
and PCA$_{\rm H\gamma_{A}}$, as well as the present-day star formation rate
over stellar mass (SSFR) derived from the extinction-corrected H$\alpha$ flux as
described in \citet{mor}, to a library of models generated from the 
Bruzual \& Charlot (2003) population synthesis code.  
As we will demonstrate, these three quantities jointly place
strong constraints on whether the galaxy has been 
forming stars smoothly or episodically over the past 1-2 Gyr.

We create a library of models by using the population synthesis code of 
\citet{bru}. The underlying model has a continuous SFR, 
declining exponentially with time SFR$(t) \propto exp(-\gamma t)$,
with $\gamma$ uniformly distributed between 0 (i.e. constant star formation rate) and  1 Gyr$^{-1}$. 
Stars begin to form at a look-back times between 13 Gyr and 1.5 Gyr. 

We then superimpose additional bursts  of star formations onto 
these continuous models. 
\footnote {It is true that exponentially decaying star formation histories are
not very physical. However, as we will show,  the look-back times
when half the stellar mass was formed is never less than $\sim$4 Gyr.  
As a result, an exponential model and other parametrizations such as a delayed
exponential model will yield the same answers regarding burst requirements, because
the timescale over which the H$\gamma_{A}$ index becomes strong following a burst
is $\sim$1.5 Gyr.}
The strength of the burst is defined as the fraction 
of the stellar mass  produced by bursts relative to the total mass formed
by continuous models, and is distributed logarithmically between 0.001 and 4. 
Bursts are described by top-hat functions and have durations between  
3$\times10^7$ to 3$\times 10^8$ years.  
The combination of Balmer absorption lines and 4000 \AA\ break strength
is not sensitive to star formation episodes that have occurred longer than about  $2$ Gyr
ago, so the bursts occur randomly between look-back times of 2 Gyr and the present.    
The metallicity range of our models is distributed uniformly 
from 0.005 to 2.5$Z_{\sun}$. Each model includes dust extinction based on 
the two-component model of \citet{cha}, where the  V-band optical 
depth $\tau_{v}$ follows a Gaussian 
distribution with a peak at 1.78 and a width of $\sigma$ = 0.55.
The $\tau_{v}$ distribution is truncated so that it only spans the range 0-4. 
$\mu$ is the  fraction of that optical depth affecting stellar populations older than 
10 Myr and is uniformly distributed between 0 and 1.

\subsection{Fitting procedure} 

We first discard those spectral bins with contamination from 
background or foreground sources (3 spectral bins in total, originating from 2 galaxies). 
We also exclude the spectral bins with large errors in the PCA indices ---
PCA$_{\rm H\gamma_{A}}>1$ and PCA$_{\rm D_n(4000)}$$>0.2$. 
This leaves a total of 2725 spectral bins 
out of the original 2812 bins. We first check whether  
the 2 PCA indices and the SSFR can be fit using the library of continuous SF models. The bins where 
such fits fail are then fit with the full library of   
continuous and burst models. We have used both the best-fit model with minimum 
chi-square to estimate quantities such as the stellar mass formed in
the past 2 Gyr, as well as the median of the probability distribution function.
In practice, the two estimates yield virtually identical results (see Figs 1 and 2). We record  
the stellar mass produced in the last 2 Gyrs. When the continuous models do not
provide acceptable fits  within 1$\sigma$ error of data, we also record the 
stellar mass produced in the burst mode during this time period.

\subsection{Separating inner and outer regions of galaxies}

In the following sections, we will often refer to the mass of recently formed 
stars in both the ``inner" and the ``outer" regions of the galaxy. 
We choose 0.7R$_{90}$ as the nominal radius to partition our galaxies 
into inner versus outer spectral bins,  
where R$_{90}$ is the radius enclosing 90 percent of the $r$-band Petrosian flux.
In \citet{mor}, they found that there was a clear anti-correlation between
the outer-disk metallicity (R $> 0.7R_{90}$) and the total galaxy H\textsc{i} fraction. 
We therefore follow their definitions of the inner and outer regions of galaxies.
We weight each spectral bin by the stellar mass enclosed by the circular annulus formed by   
the inner and outer boundaries of the bin.  
We then calculate weighted  averages of quantities such as 
the fraction of recently-formed stellar mass and gas-phase metallicity.

\subsection{Categorization of Galaxies According to Their Star Formation Histories}

We divide our galaxy sample into five classes according to their star formation
histories in their inner and outer regions: 

(1)Type A: Those that are well described by continuous models at all radii, i.e.
the star formation histories are well described by a varying exponential decline 
time as a function of radius.  
Type A galaxies have t$_{start}$ ranging between 13 and 12 Gyr in look-back time.  

(2)Type B: Those that are well described by continuous models at all radii, but 
with t$_{start}$  varying between 12 and 2 Gyr. 

(3)Type C: Those that are well described by continuous models with
 t$_{start}$  varying between 13 and 2 Gyr in their inner regions, but which require a   
burst in their outer regions.  

(4)Type D: Those that require bursts in both their inner and their outer regions.  

(5)Type E: Those that require bursts only in their inner regions.

\begin{figure}
 \includegraphics[width= 0.45\textwidth]{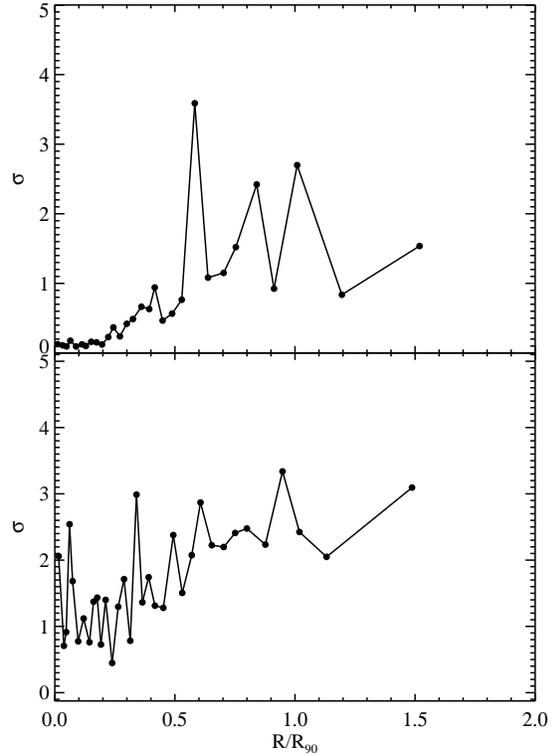}
    \caption{The average deviation of the observed values of D$_n$(4000), H$\gamma_A$
and SSFR from the values predicted by the best-fit continuous models for all Type C (upper panel) 
and Type D (lower panel) galaxies as a function of radius. Note that the deviations are  scaled
by the  measurement errors $\sigma$ and are  plotted in units of $\sigma$. This takes into
account the fact that the $S/N$ of the binned data is higher in the inner regions
than the outer regions. }
   \label{sigma}
\end{figure}

The number of galaxies of type A, B, C, D, and E is 45, 24, 34, 104, 2 respectively. 
Some examples of each type are presented in Figs \ref{typeAB} and \ref{typeCD}. Fig.
\ref{typeAB} shows examples of galaxies of types A and B, which are well fit 
by pure exponential models. In the first row, we  show SDSS cut-out images.
The 2nd, 3rd, 4th rows show  SSFR, D4000 and H$\gamma_{A}$  as a function of radius in the galaxy. 
Note that the radii have been scaled by dividing by R$_{90}$. In these panels, black points show the
data along with the errors. Red/blue lines are model fits using the     
minimum chi-square models and the median of the PDF, respectively. 
In the 5th row, we plot the parameters of the minimum chi-square models 
as a function of radius. The 5th row shows the look-back time when half 
of stellar mass was formed at each radius.

Fig. \ref{typeCD} shows examples of galaxies of types C and D, which require bursts. The rows
are the same as in the previous figure. In the left column, we show the best-fit 
continuous models. As can be seen, in both cases such models generally work better 
near the centers of galaxies. 
In the outer regions, the deviations in the fit to PCA$_{\rm H\gamma_{A}}$ and 
PCA$_{\rm D_n(4000)}$ become large. The right panels show how addition of
a burst bring the models into much better agreement with the data.
We note that the physical size of each bin for our sample is at 
least $\sim$1.5-3kpc, and the discrepancies between the data and the 
continuous models often span more than 2 bins. This means that the bursts
are not generated by individual H\textsc{ii} regions, but span a significant 
portion of the outer disk.

In Fig. \ref{sigma}
we plot the {\em average} deviation of the continuous models from the observations as a function of radius for 
type C (upper panel) and type D (lower panel) galaxies. In order to account for
the fact that the observational errors increase as function of radius, we
scale the  deviation by $\sigma$ before taking the average . As expected,   
the centers of Type C galaxies are well fit, with the average deviation increasing towards
the outer regions.  
The lower panel of 
Fig. \ref{sigma} shows that the same general trend is also true for type D galaxies.
In the central regions, the average deviation from a continuous model is only
$\sim 1 \sigma$, but in the outer regions the average deviation increases by a factor of 2-3.
As we will show in the next section, the main difference between galaxies with type C and
type D star formation histories is stellar mass. Galaxies with recent star formation only
in their outer regions are predominantly systems with stellar masses greater than $\sim 10^{11} M_{\odot}$.

\begin{figure*} 
  \begin{center}
    \subfigure
    {\includegraphics[width=0.497\textwidth, angle=360]{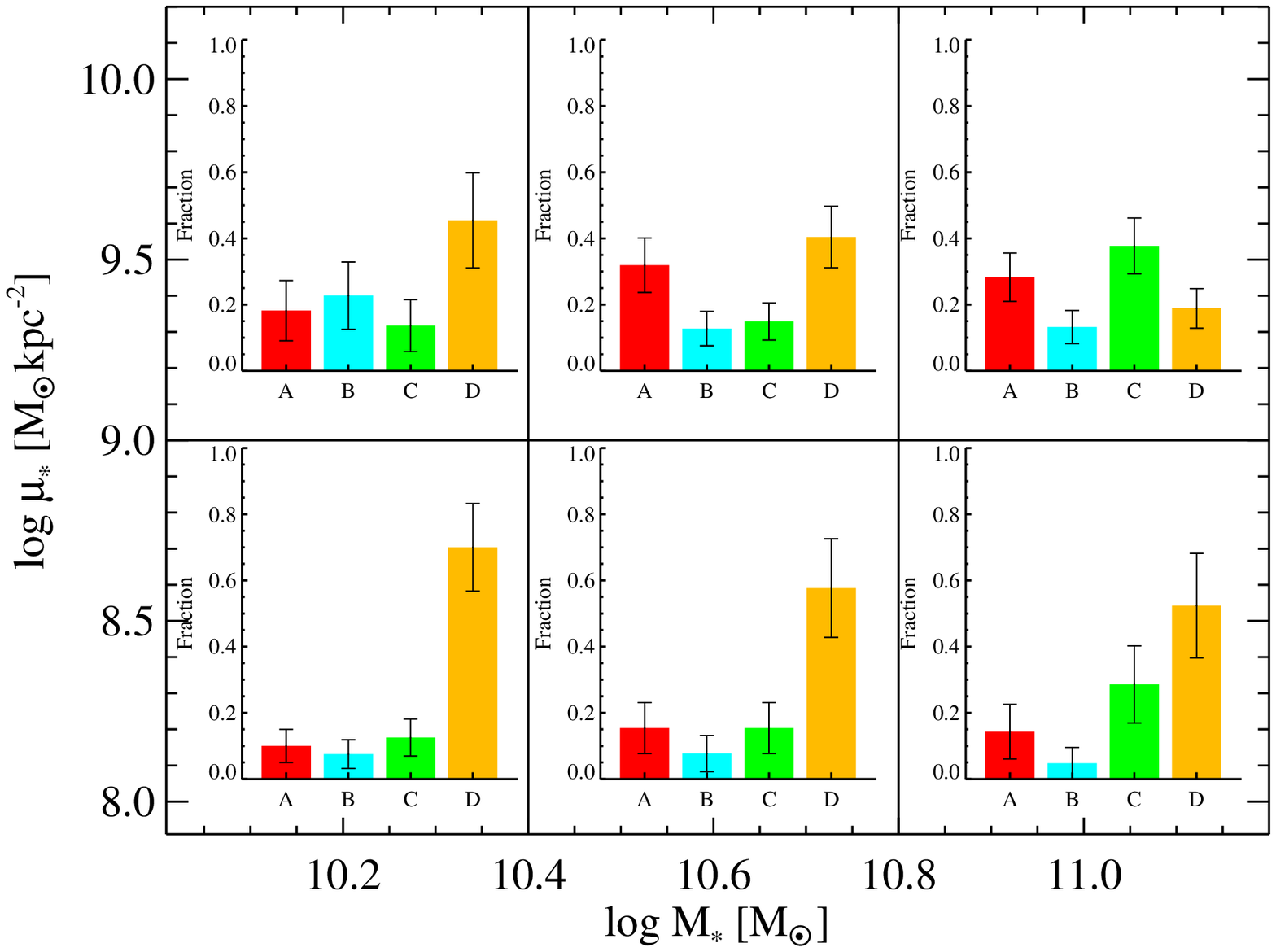}}
    \subfigure
    {\includegraphics[width=0.497\textwidth, angle=360]{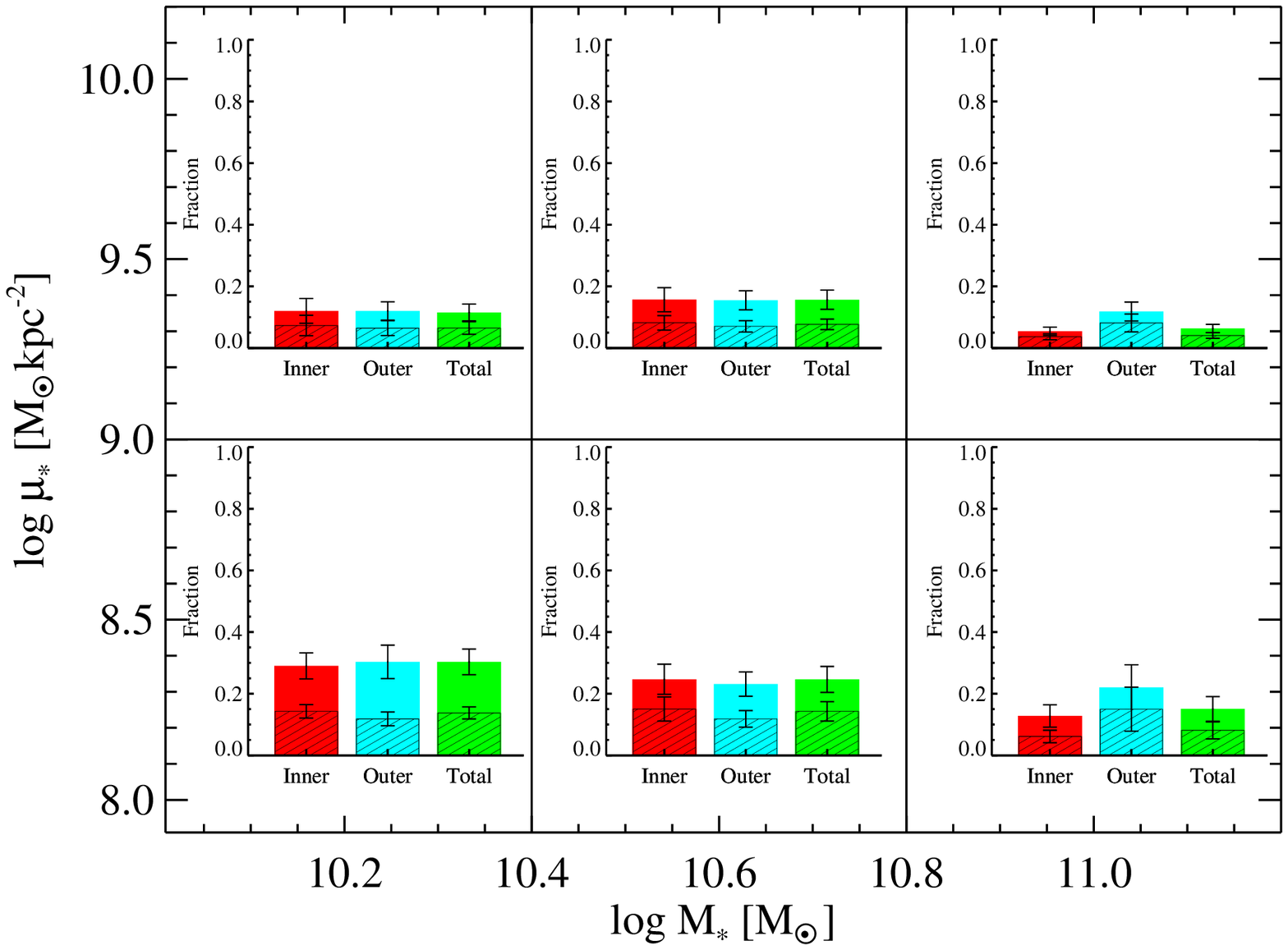}}
\caption{Left: Fraction of galaxies with different types of star formation history
in the two-dimensional plane of stellar surface density versus stellar mass. 
Poisson error bars are shown. Right: Fraction of the 
stellar mass formed in the last 2 Gyr in the inner region ($R< 0.7 R_{90}$, red), 
outer region ( $R> 0.7 R_{90}$, cyan), and
in the whole galaxy (green). The hatched area shows the fraction of 
stellar mass formed in the last 2 Gyr in the {\em burst mode}. Error bars have been computed using
boot-strap resampling.}
\label{smd}
  \end{center}
\end{figure*}

\begin{figure*} 
  \begin{center}
    \subfigure
    {\includegraphics[width=0.497\textwidth, angle=360]{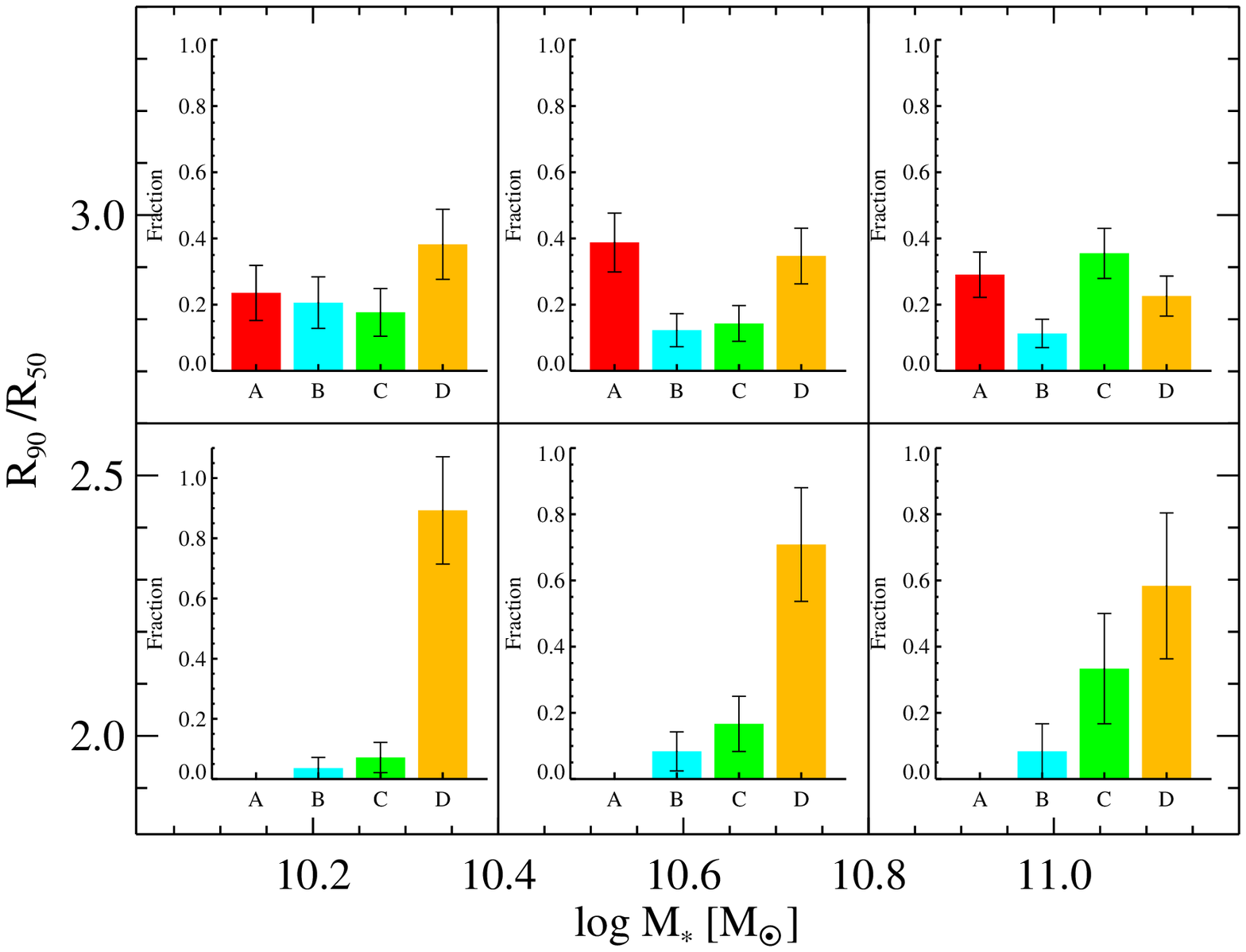}}
    \subfigure
    {\includegraphics[width=0.497\textwidth, angle=360]{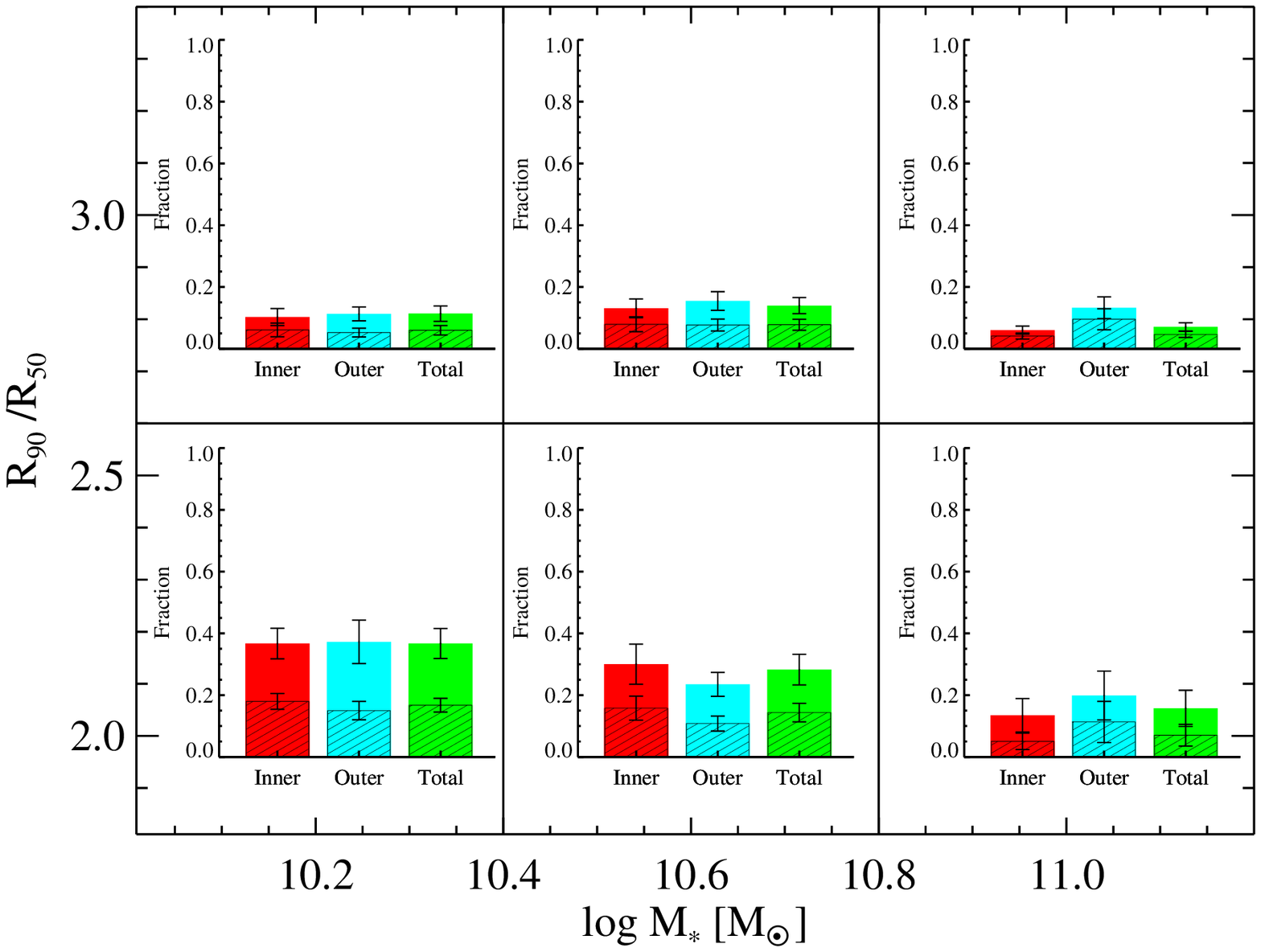}}
\caption{As in Fig.\ref{smd}, except for galaxies in the two dimensional plane of
concentration versus stellar mass.}
\label{cix}
  \end{center}
\end{figure*}

\begin{figure*} 
  \begin{center}
    \subfigure
    {\includegraphics[width=0.497\textwidth, angle=360]{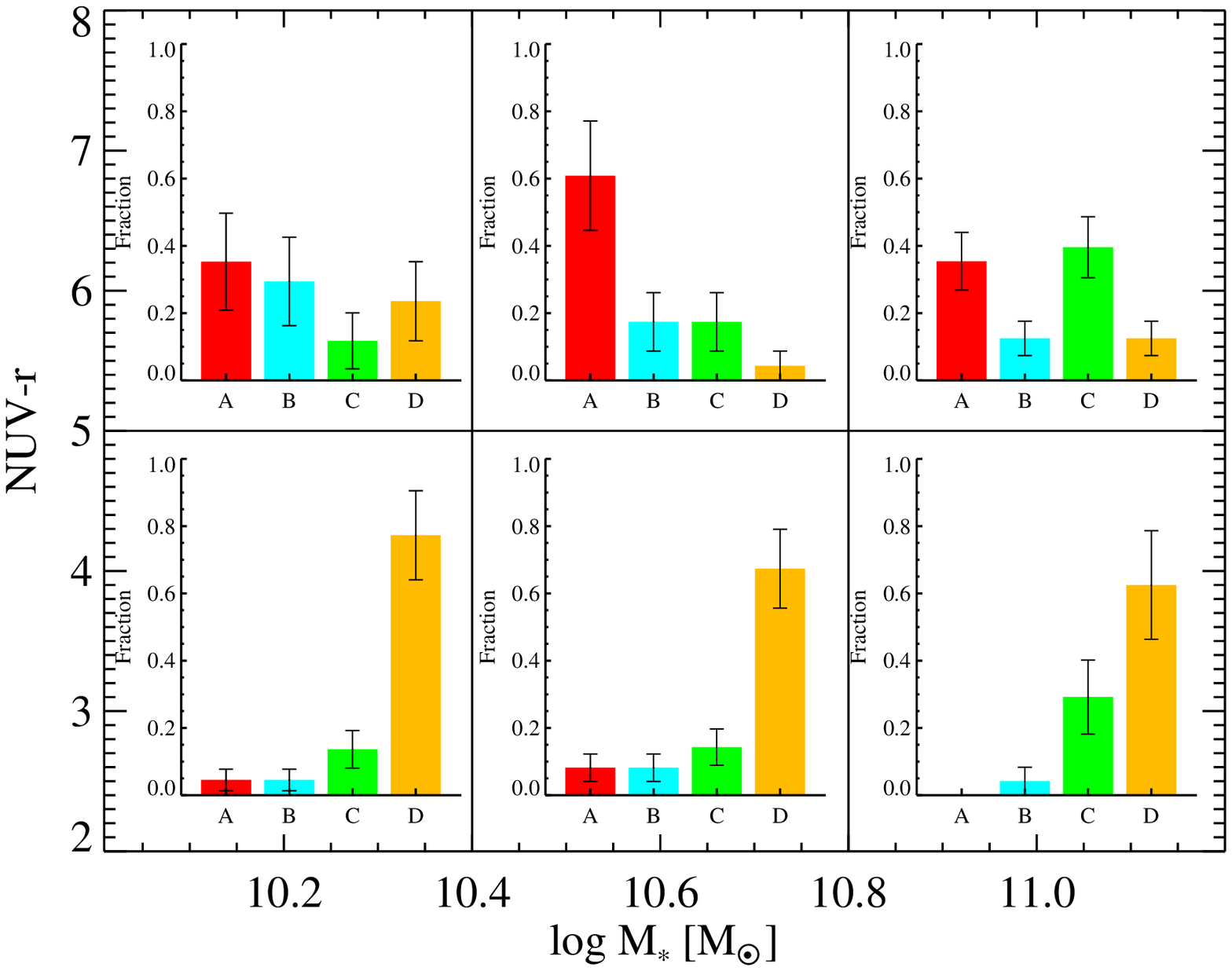}}
    \subfigure
    {\includegraphics[width=0.497\textwidth, angle=360]{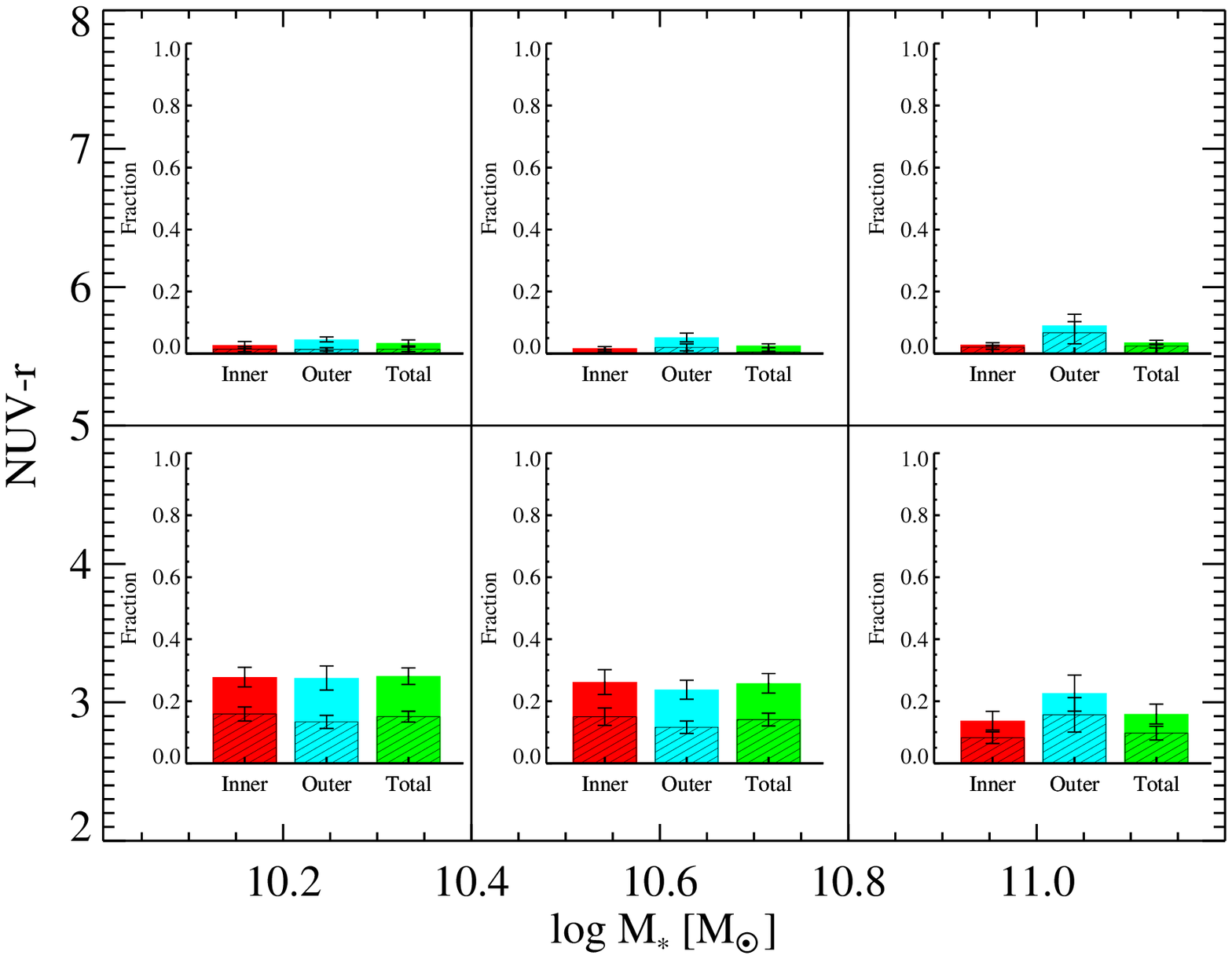}}
\caption{As in Fig. \ref{smd}, except for galaxies in the two dimensional plane of
NUV-$r$ colour versus stellar mass.}
\label{clr}
  \end{center}
\end{figure*}

\section[]{Results }

\subsection{ Dependence of star formation history on stellar mass, surface density, concentration
and colour}

In this section, we investigate how the fraction of galaxies with different
types of star formation histories defined in Section 3.5 depends on stellar mass, galaxy 
colour and structural parameters such as surface density and concentration. 

For central galaxies, stellar mass is a good tracer of the mass of dark matter halo in which the
galaxy resides. The surface density of the galaxy is a rough measure of the degree to which
the gas has lost angular momentum prior to being converted into stars.
The concentration index is a good proxy for the bulge-to-disk or 
bulge-to-total luminosity ratio of the galaxy (see Fig. 1 of \citealt{wei}). 
Finally, in the absence of dust,  the global NUV-r colour is a measure of the average luminosity-weighted age 
of the stellar population.
In order to understand whether the star formation history of a galaxy is
determined primarily by its mass or by its structural properties, we will analyze SFH trends 
in {\em two dimensional planes} of stellar mass versus surface density and concentration.
We will also examine how the fraction of stars formed both continuously 
and in bursts in the inner and outer regions of the galaxy
depends on the location of the galaxies in these 2D planes.

\subsubsection{The plane of stellar mass versus stellar surface density}

All galaxies are divided into 6 bins by dividing their stellar mass into 3 bins, 
10.0--10.4 $M_{\odot}$, 10.4--10.8 $M_{\odot}$, and 10.8--11.2 $M_{\odot}$ in $\log M_*$, 
and surface mass density into 2 bins: 
$\log \mu_* < 9$ $M_{\odot}$ kpc$^{-2}$ and $\log \mu_* > 9$ $M_{\odot}$ kpc$^{-2}$ . 
In each bin, we calculate (1) the fraction of galaxies with different 
types of star formation histories (left panel ), (2) the fraction of the stellar mass 
formed within the last 2 Gyr in the inner region, the outer region and in the entire  
galaxies (right panel), (3) the fraction of the stellar mass formed in the {\em burst mode} 
in the last 2 Gyr in the inner region, the outer region, and 
in the entire galaxy (hashed histograms in the right panel).  

As shown in the left panel of Fig. \ref{smd}, the star formation histories of galaxies
clearly depend on both stellar mass and surface density. 
The fraction of quiescent galaxies without bursts (types A and B) depends mainly on 
surface density rather than stellar mass. There is some tendency for quiescent galaxies    
to be somewhat more active (i.e. be of type B) if they have low masses.

The fraction of galaxies that  have experienced recent bursts in both their inner
and their outer regions (type D) decreases both as a function of stellar mass 
and stellar surface density. In the stellar mass bin with $\log M_*$ in the 
range 10--10.4, type Ds make up more than two thirds of the entire population, 
but type Ds make up only a small fraction of the galaxies our highest $M_*$/$\mu_*$ bin.
Interestingly, however,  the fraction of galaxies that have experienced recent bursts in
their outer regions only (type Cs) exhibits the opposite dependence of
on the mass of the galaxy. 
In the highest stellar mass bin with $\log M_*$ in the range
10.8--11.2, there are as many type C galaxies as there are quiescent (type A and B) galaxies.  

The right panel of Fig. \ref{smd} displays the fraction of stars formed in the last 2 Gyr
as a function of location in the plane of stellar mass and stellar surface density.
We show the fraction of recently formed stars in the whole galaxy (in green)
as well as the fraction in the inner ($R< 0.7 R_{90}$, red) and outer ($R>0.7 R_{90}$,blue)  
regions. The fraction of stars formed in the last 2 Gyr in the {\em burst mode} is indicated
as hashed shading. As expected, the fraction of recently formed stars is highest for galaxies 
with low masses and low surface densities and lowest for galaxies with high masses and 
high surface densities. Interestingly, the recent star formation is always rather evenly split  
between the inner and outer parts of galaxies, except for the most  massive
galaxies with low stellar surface densities, where more star formation occurs in the galaxy 
outskirts. These results are consistent with a recent analysis by \citet{per} of star 
formation in the inner and outer regions of galaxies using a sample of $\sim 100$ galaxies 
with integral field unit (IFU) spectroscopy. What is new here, is that we can estimate 
the fraction of the recent star formation that occurred in bursts.
We see that the burst fraction is highest in  massive galaxies.  The majority of galaxies 
with stellar masses in the range 10--10.4 in $\log M_*$ require bursts in both their inner 
and in their outer regions (i.e. they are classified as type D systems), but the total mass of 
stars formed in the bursts over the last 2 Gyr is comparable or smaller (in the case of
low mass galaxies with low densities) than that formed in the quiescent
mode. In high mass galaxies, the opposite is true -- the majority of the recent 
star formation is contributed by bursts.

\subsubsection{The plane of stellar mass versus concentration }

Fig. \ref{cix} is similar to Fig. \ref{smd} except that we divide galaxies
according to concentration index ($R_{90}/R_{50}$) at a value of 2.6.  
As can be seen, the results are quite similar to those obtained
when galaxies are divided by stellar surface mass density.
One notable difference is that in Fig. \ref{cix} we see that {\em all} type A 
galaxies have the division between quiescent galaxies and galaxies
with actively ongoing star formation is much more pronounced. 
 We see that {\em all} type A galaxies have $R_{90}/R_{50} > 2.6$, 
whereas Fig. \ref{smd} shows that there is a significant number of type A galaxies 
with stellar surface mass densities less than $10^9 M_{\odot}$ kpc$^{-2}$. 
We note that high concentration index (equivalently bulge-to-disk ratio) instead of high
stellar surface density (equivalently contraction factor) appears to be the necessary
condition for galaxies with quiescent type A star formation histories.
Although high concentration is a necessary condition, it is not a sufficient 
condition -- a sizeable fraction of bulge-dominated galaxies have experienced 
bursts of star-formation in the last 2 Gyr.

\subsubsection{The plane of stellar mass versus colour }
In Fig. \ref{clr} we divide galaxies according to 
(NUV-r) colour at a value of 4.8 . As can be seen from the right-hand panel,
red global colours select out the galaxy population that have experienced very
little recent star formation both in their inner and in their outer regions.
It is noteworthy that almost all galaxies with quiescent star formation
histories (types A and B) belong to the red population with NUV-r $> 4.8$. 
Galaxies that have experienced bursts in their outer regions are roughly
split evenly between the red and the blue populations, while type D
galaxies that have experienced bursts in both their inner and outer
regions are predominantly blue. We note, however, that a significant population of red
type D galaxies is found in the lowest stellar mass bin.

\begin{figure*} 
 \begin{center}
  \subfigure
  {\includegraphics[width=0.65\textwidth, angle=360]{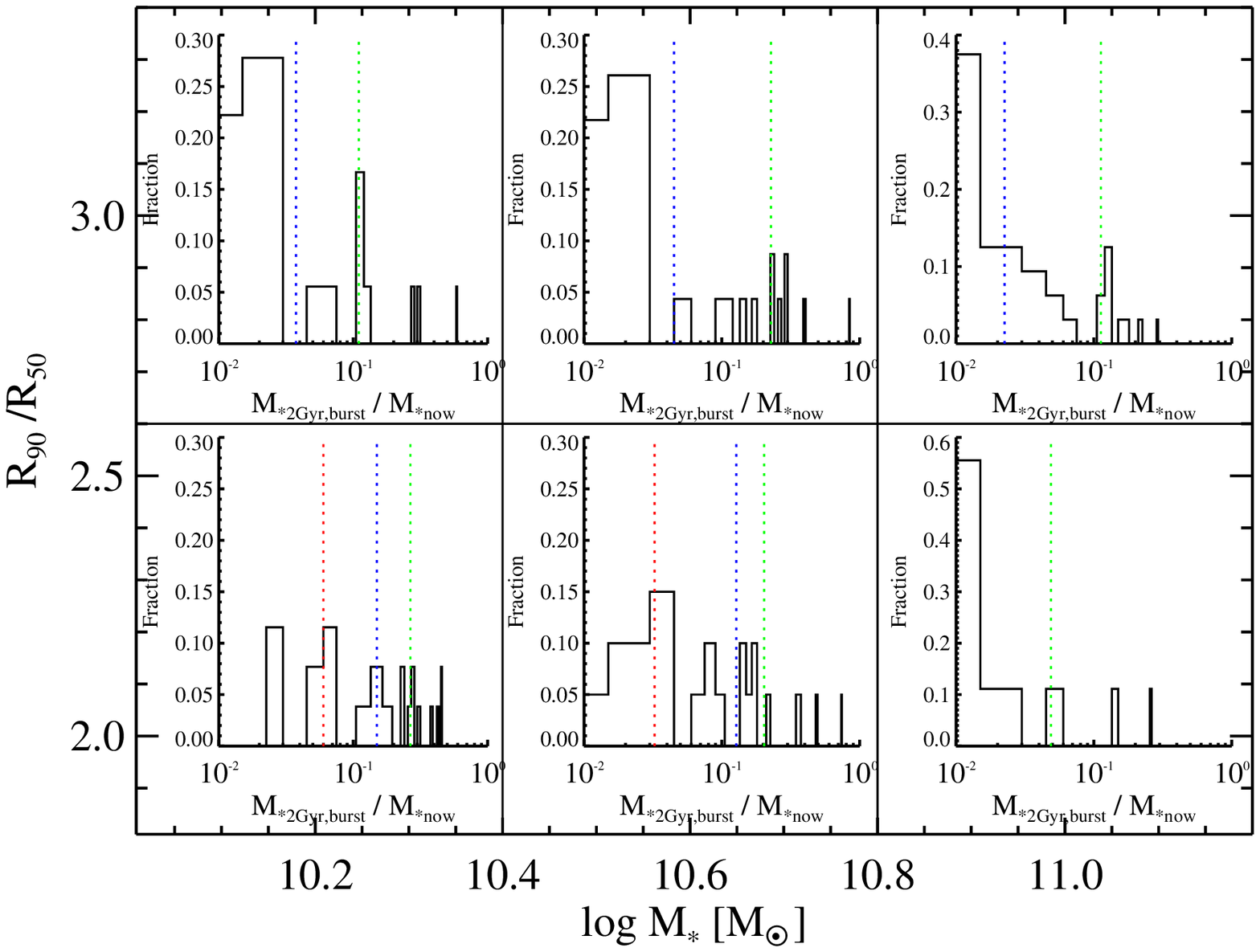}}
  \subfigure
  {\includegraphics[width=0.65\textwidth, angle=360]{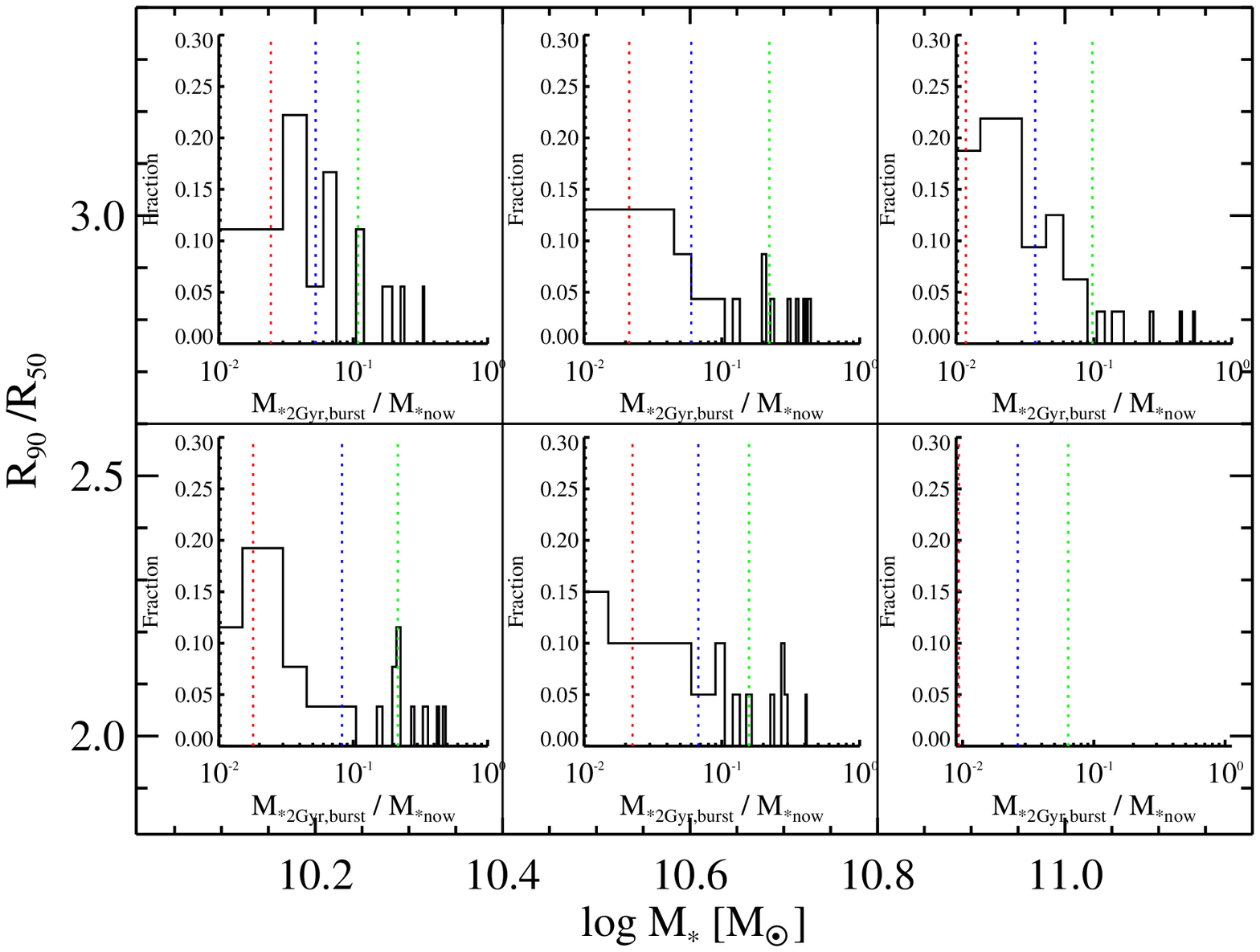}}
\caption{Top: Distribution of the fraction of stars formed in bursts in the inner 
         regions of galaxies in the two-dimensional plane of concentration versus 
         stellar mass. Bottom: As in the top panel, but in the outer region. 
         The red, blue, green dotted lines indicate 25\%, median, 75\% values of  
         the distribution in each panel. } 
\label{f2gbst}
 \end{center}
\end{figure*}

\begin{figure*} 
 \begin{center}
  \subfigure
  {\includegraphics[width=0.65\textwidth, angle=360]{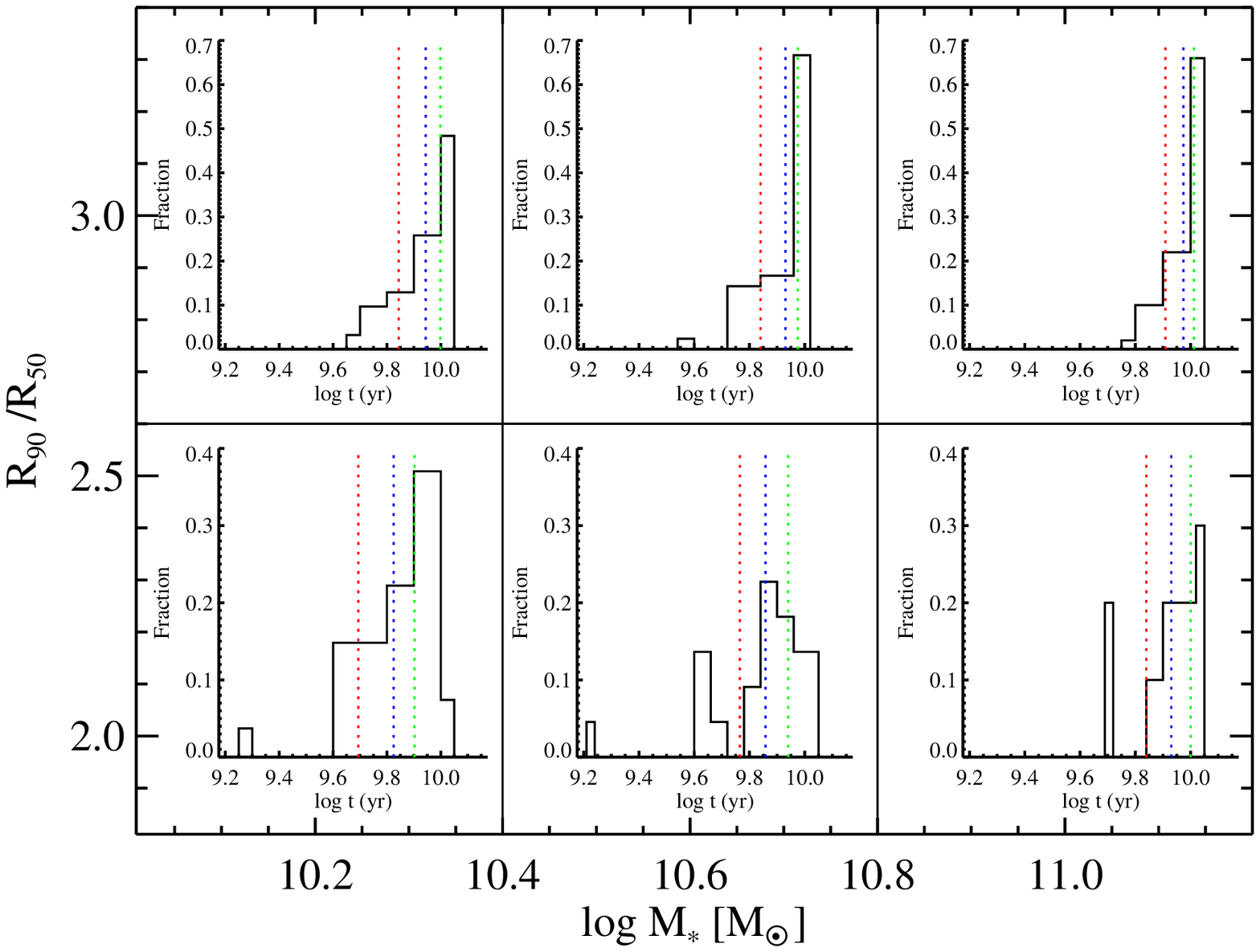}}
  \subfigure
  {\includegraphics[width=0.65\textwidth, angle=360]{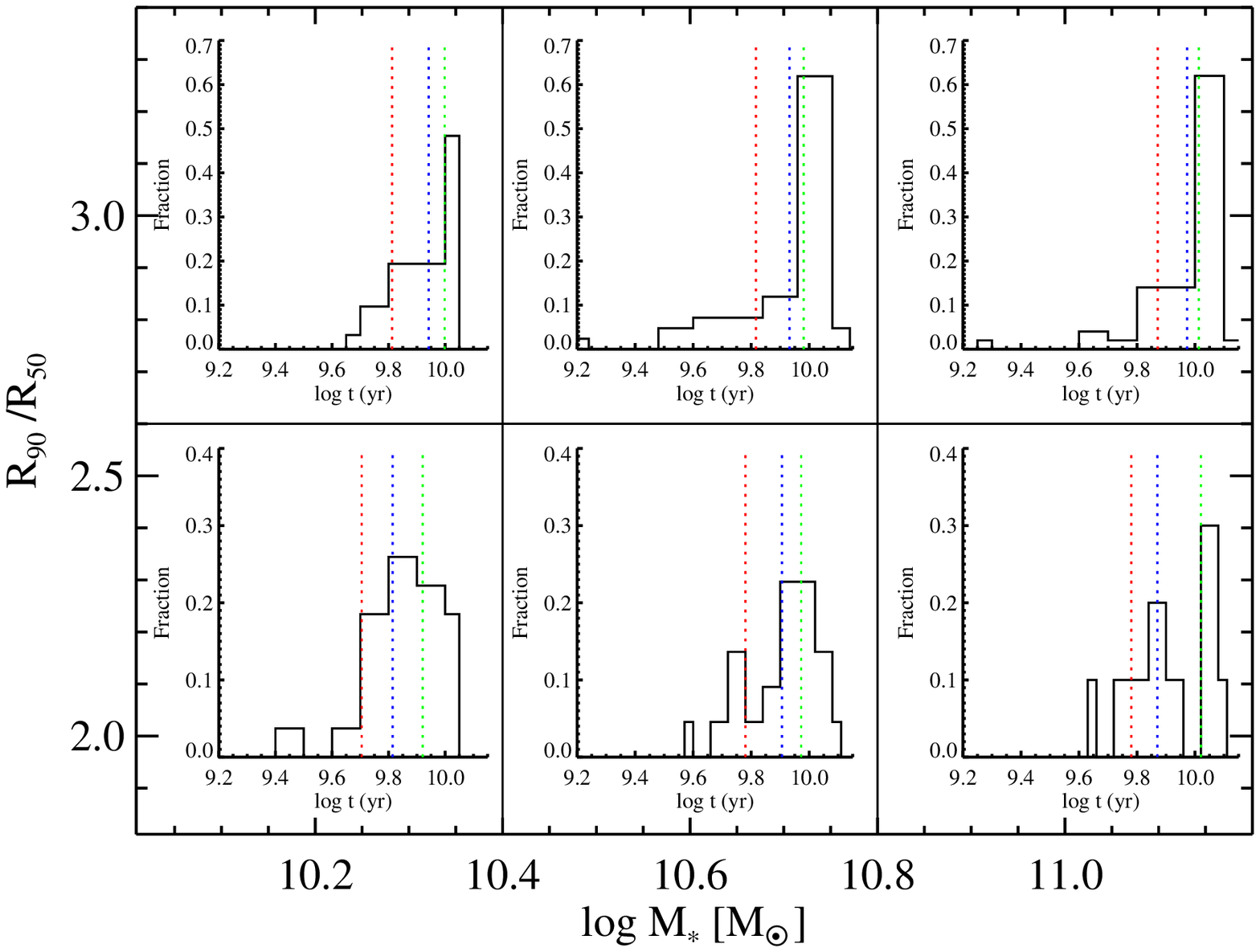}}
\caption{Top: Distribution of the look-back time when half the stellar mass was formed
         in the inner regions of galaxies in the two-dimensional plane of concentration
         versus stellar mass. Bottom: As in the top panel, but in the outer region.
         The red, blue, green dotted lines indicate 25\%, median, 75\% values of  
         the distribution in each panel.} 
\label{t50}
 \end{center}
\end{figure*}

\subsubsection{Distribution of SFH parameters of galaxies}

In the previous sub-sections we studied the fraction of galaxies with
bursty versus continuous star formation histories as a function of stellar mass, 
colour and structural
parameters. We also looked at the contribution of bursts to the recent star formation in
the inner and the outer regions of galaxies, as a function of these same parameters.
In this sub-section, we analyze the distribution of burst strengths in galaxies, as
well as the look-back time when half the stars were formed. For simplicity, we 
confine our attention to the two dimensional plane of  concentration
versus stellar mass. 

(i) {\it The Fraction of stars formed in bursts.} Fig. \ref{f2gbst} shows the 
distribution of the fraction of stars formed in bursts 
in the inner and outer regions of galaxies as a function of
position within this plane . As seen in the 
top panel,  the fractions of stars formed in bursts in the inner region is generally
around a few percent for bulgy galaxies ($R_{90}/R_{50} > 2.6$) or for massive galaxies 
($\log M_* > 10.8$). In the inner regions of less massive, disk-dominated galaxies 
the median burst strength is $\sim$10\%  and the distribution is much wider.  
The bottom panel of Fig. \ref{f2gbst} shows that burst strengths in the outer regions of
galaxies are insensitive to whether it is bulge or disk-dominated. The main controlling
factor is the mass of the galaxy. In the highest stellar mass bin, the
median burst strength in the outer region is $\sim3$\%,
about a factor of two lower than in the other two bins.  
We also note that in massive galaxies, the outer bursts are stronger than the inner bursts
on average, whereas the opposite is true for low mass galaxies.

(ii){\it The look-back time when half the stellar mass was formed}. In Fig. \ref{t50}, 
we make similar plots of the look-back time when half the stellar mass was 
formed (T$_{50}$), which provides 
information about the mean stellar ages of the galaxies. The trends in this plot 
are much weaker. Only a very small fraction of galaxies have T$_{50}$ less 
than $\sim 4 \times 10^9$ Gyr in either their inner or their outer regions. 
This is in keeping with the results in the previous section, which shows that although 
recent bursts are required to fit the spectral parameters, the fraction of mass formed
in these bursts generally does not exceed $\sim 10-20 \%$ of the total stellar mass.   
Disk-dominated galaxies 
generally have younger mean stellar ages than bulge-dominated galaxies
of the same stellar mass,
and  the most massive galaxies have the oldest mean stellar ages. 
The trend for the  most massive, bulge-dominated galaxies
to have the oldest mean stellar ages holds in both the inner and outer regions of the galaxies, but
is somewhat stronger in the outer region.

We also note that outer star formation in massive galaxies 
discussed in the previous section, is manifested in this plot as shift in the
lower 25th percentile of the mean age distribution, rather than a shift in the median
value. This means that outer disk growth is a phenomenon that pertains to a {\em minority
of the massive galaxies}, rather than to the population in general.

\begin{figure} 
 \begin{center}
  \subfigure
  {\includegraphics[width=0.45\textwidth, angle=360]{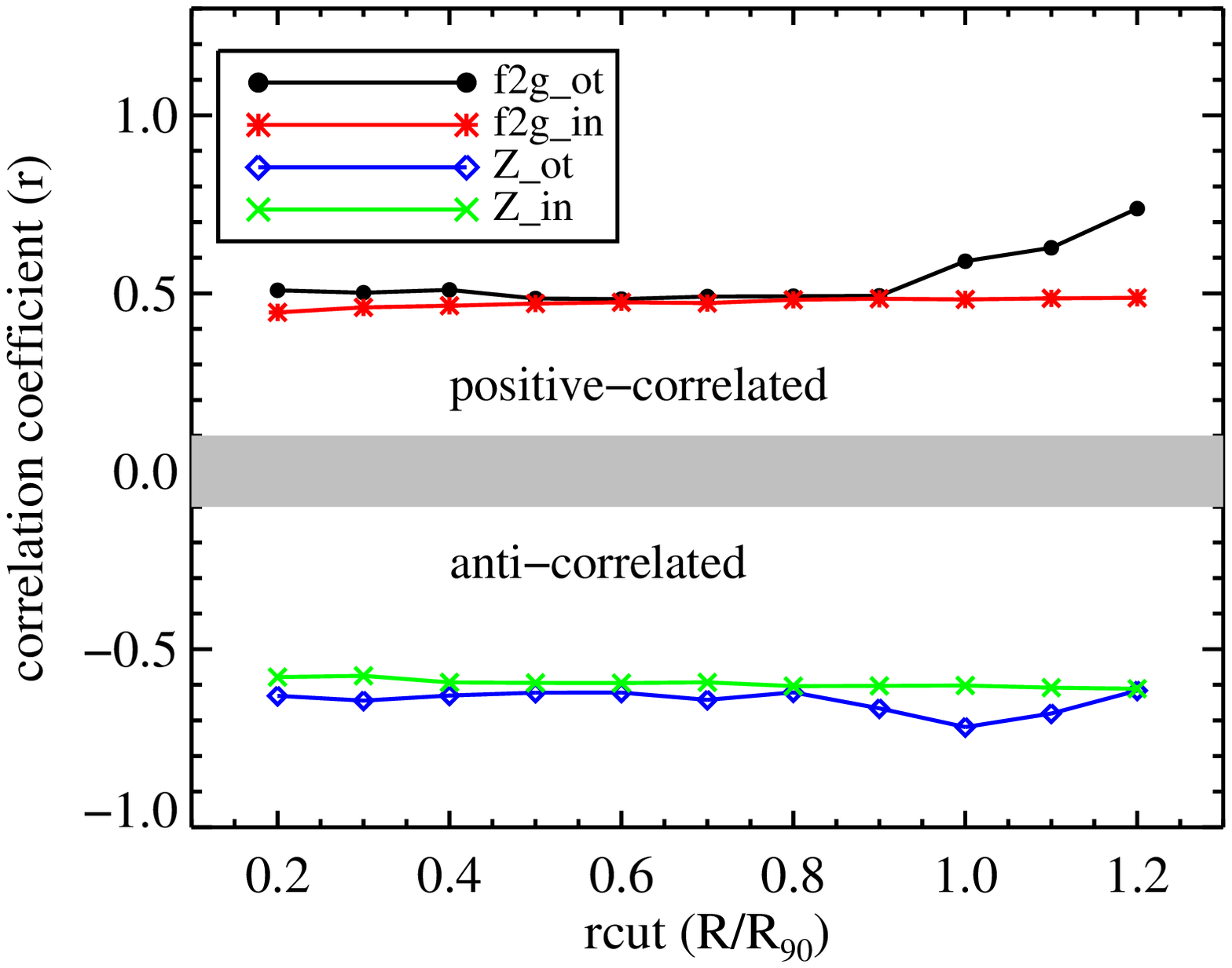}}
  \subfigure
  {\includegraphics[width=0.45\textwidth, angle=360]{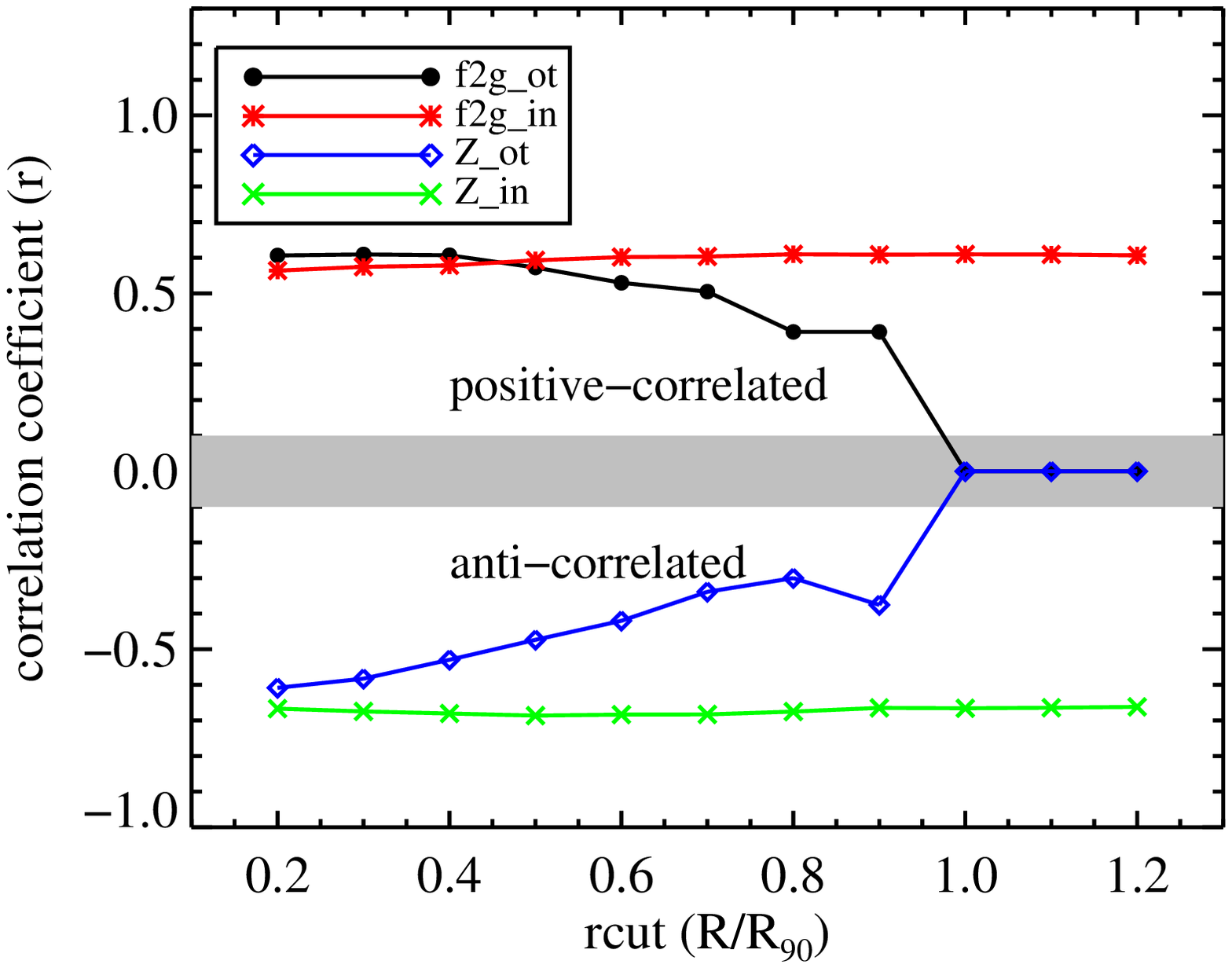}}
\caption{Top: The correlation coefficients of Spearman rank tests for the 
fraction of recently-formed stellar mass and gas-phase metallicity  as a function of HI mass fraction. 
The four parameters,f2g\_in, f2g\_ot, Z\_in, Z\_ot, are M$_{*2Gyr}/M_{*now}$ for the
inner region, M$_{*2Gyr}/M_{*now}$ for  the outer region, inner metallicity, and outer 
metallicity. The correlation coefficients are plotted as a function
of $R_{cut}$, the radius (in units of $R_{90}$) used to separate the inner
region of the galaxy from the outer region. Bottom: As in the top panel, but for the
four parameters as a function of H$_2$ mass fraction.} 
\label{gas}
 \end{center}
\end{figure}

\subsection{Relations between star formation history, HI/H$_2$ mass fraction and gas-phase metallicity}
In the previous section, we explored how the star formation histories in the inner and outer regions
of galaxies depend on their stellar mass, structural properties and colour.
In this section, we examine how star formation history depends on the gas content
and gas-phase metallicity of a galaxy. In previous work,
\citet{mor} showed that around 10\% of all galaxies with stellar 
masses greater than $\log M_* > 10$  exhibit strong drops in
gas-phase metallicity in their outer regions. These galaxies tended to have high H\textsc{i} content
and to have actively star-forming outer disks.   

In this analysis, we 
explore the correlations between the fraction of recently-formed stars and gas-phase metallicity 
in the inner and outer regions of the galaxy with both the H\textsc{i} and H$_2$ gas mass fraction. 
We also explore how the correlations change as we vary the radius $R_{cut}$ used to
divide the inner galaxy from the outer galaxy.

The spectral bins with only upper or lower limits on metallicity are discarded before we carry out 
our  analysis. For each value of $R_{cut}$ (in units of R$_{90}$), we compute
correlation coefficients of Spearman rank tests for
\begin {itemize}
\item The relations between the fraction of stars formed in the inner/outer regions
of the galaxy and total H\textsc{i} gas mass fraction.  
\item The relations between the fraction of stars formed in the inner/outer regions
of the galaxy and total H$_2$ gas mass fraction.  
\end {itemize}
We only take into account the correlation coefficients with probability larger than 0.97.
For the coefficients with 
probability smaller than 0.97, we set the coefficients to zero. 
(In practice, this only happens for the relations between  outer star formation and metallicity 
and H$_2$ gas mass fraction when $R_{cut}$ is greater than $R_{90}$.) 

Fig. \ref{gas} summarizes the results of this exercise.  
$rcut$ is in units of R/R$_{90}$. The upper and lower panels show the results 
for the correlations with H\textsc{i} and H$_2$ gas mass fraction, respectively.
Coefficients with values larger and smaller than zero indicate positive and anti-correlations 
between the parameters.

As can be seen in the upper panel of the figure, the fraction of recently-formed 
stars and the gas-phase metallicity
in both the inner and the outer regions of the galaxy correlate strongly with the
global atomic gas mass fraction. The strength of the correlations 
does not depend on the adopted value of $R_{cut}$, until $R_{cut}$ reaches values
greater than 0.9 R$_{90}$. At this point, stronger correlations between the gas-phase metallicity and
the fraction 
of stars formed in the outer regions and the  total H\textsc{i} content are seen.
Interestingly, the correlation between the fraction of stars formed in the inner region of the galaxy
and total H\textsc{i} gas fraction is  independent of the value of  $R_{cut}$.
Even as $R_{cut}$ decreases  to values near zero,  there is still a correlation between
the fraction of stars formed in the very inner regions of the galaxy and its total atomic gas content. 
We will discuss possible reasons for this in the next section.  

In the lower panel of Fig. \ref{gas}, we see that the gas-phase metallicity and
the fraction of recently-formed stars in the inner regions of the galaxy are also both correlated
with H$_2$ mass fraction. The correlation of these inner quantities with
H$_2$ mass fraction is stronger than the correlation with H\textsc{i} mass fraction
($r$ values of $\sim0.6$ rather than $\sim0.4$). Similar to what is seen for the atomic gas, the 
strength of the correlations of the inner quantities with H$_2$ gas mass fraction are independent of $R_{cut}$. 
The outer quantities, however, are much more weakly  
correlated with H$_{2}$ mass fraction and for  $R_{cut}> 0.7$, the correlations
with H$_{2}$ mass fraction disappear entirely.

\section{summary and discussion} 

We now summarize our main results as follows:
 
(i) By fitting stellar population models to the combination of SSFR, D4000, H$\gamma_{A}$, 
we show that the star formation histories of many disk galaxies cannot be accurately
represented if  the star formation rate has declined exponentially as a function of time.
Many galaxies have strong Balmer absorption lines that require recent short-lived episodes 
of star formation in both their inner and in their outer regions. 
 
(ii) The fraction of galaxies that have experienced such episodes in both
their inner and outer regions is highest for systems with low stellar masses, 
low surface densities, and low concentrations, i.e., late-type galaxies. 

(iii) Around a third of all massive ($\log M_* > 10.8$), bulge-dominated galaxies
have experienced recent star formation episodes only in their {\em outer} regions.

(iv) For low mass, disk-dominated galaxies, the fraction of stars formed in a single burst   
episode is typically around 15\% of the total stellar mass in the inner regions of the galaxy 
and around 5\% in the outer regions of the galaxy. When we average over the
population, we find that such bursts contribute around a half of  total mass in stars
formed in the last 2 Gyr. 

(v) For massive galaxies ($\log M_* > 10.8$), the fraction of stars formed in bursts
is only $\sim 2-3$ \% . Averaging over the population, however, we find that such bursts contributed 
nearly all the mass in stars formed in the last 2 Gyr.  

(vi) The amount of recent star formation in both the inner and outer regions of a galaxy is
positively-correlated with its total atomic gas content. In contrast, only the inner star formation
is correlated with total molecular gas content. 

(vii) Very similar results are obtained for gas-phase metallicity. Metallicity 
in both the inner and outer regions of galaxies
are negatively correlated with global atomic gas fractions. The metallicity in the inner
region of the galaxy is negatively correlated with the molecular gas mass fraction.

We hypothesize that these results can be understood if galaxies accrete
atomic gas and form stars episodically. The fact that episodic component of the star formation
occurs primarily in the outer regions of massive galaxies (right panels of Figs.
\ref{smd} and \ref{cix})  argues that gas may be  accreted
at large radii. We note that the non-episodic (continuous) component of the star formation
is much more evenly split between the inner and the outer regions in these systems.

In low mass galaxies, both the episodic and the continuous components of the star formation
are evenly divided between the inner and outer regions of the galaxy. It is tempting to postulate
that this may indicate that gas accretion occurs in a different mode in low
mass systems -- some theoretical models predict that gas should be accreting
in the form of ``cold streams'' in dark matter halos with low mass, and from
quasi-static atmospheres of hot gas in dark matter halos of high mass (e.g.
\citealt{ker}; \citealt{dek}). 

However, our data does not allow us to 
rule out the possibility that the inner and outer bursts in galaxies may 
have different triggers, for example gas accretion may be the trigger in the outer regions and
disk-driven instabilities in the inner regions. Indeed, at fixed stellar mass, 
the fact that burst strengths are larger in the interiors of disk-dominated galaxies                      
densities (Fig. \ref{f2gbst}, top panel), argues that disk instabilities may indeed be key
to the origin of the episodic nature of the inner star formation. The Toomre Q
parameter scales inversely with disk surface density.  

The idea that gas accretion at large radii is followed by instability-triggered radial
inflows of gas and a burst of star formation in the central regions of the galaxy is also
consistent with our result that gas-phase metallicity and the fraction of recently formed
stars in the inner region of the galaxy is best correlated with its molecular gas
content. Molecular gas extends to much smaller radii than the atomic gas in all disk galaxies
and also exists at considerably higher surface densities \citep{ler}.
Why then do metallicity and recent star formation in the inner regions of galaxies
also correlate (albeit more weakly) with the total atomic gas content of the galaxy?
The most reasonable explanation, in our view, is that the atomic gas represents
a longer-term {\em reservoir} for star formation in the inner galaxy. In other words,
if the galaxy has no atomic gas, there is simply no raw material to be transported inwards to
be converted into molecular gas and stars at galactic centers. 
 
Finally, we would like to note that we began this paper with the 
statement ``It is now well established that galaxies in the nearby Universe separate
rather cleanly into two classes: those with disky morphologies, plentiful
gas and ongoing star formation and those that are bulge-dominated, with
little gas and star formation, and where star formation has largely ceased.''
Examination of the left panel of Fig. \ref{smd} indicates that even at high
stellar masses,  bulge-dominated galaxies
have actually experienced a wide variety of different star formation
histories. Those that are ``dead'' in that they have not experienced any recent star formation episodes
comprise less than half of the total population. In many of the massive, bulge-dominated galaxies
star formation occurs  in the outer regions of the galaxy, where the light has simply not 
been picked up by large scale redshift surveys that employ  single fibre spectrographs. 
The fibres typically have diameters of 2-3 arcseconds and only sample
light from the inner regions of the galaxies. 

The realization that many early-type galaxies
do have extended star-forming disks first came from studies of early-type galaxies
at ultra-violet wavelengths (e.g., \citealt{yi}; \citealt{kau07}; \citealt{fan}),
and these results are now being confirmed by IFU studies of the nearby galaxy population
\citet{per}.
Our study reveals that the star formation in these disks did not occur continuously over a Hubble time,
but was concentrated in a recent episode or burst. 
In future, large-scale Integral Field Unit (IFU) surveys   
of galaxies (see for example \citealt{san}) will shed more light on the nature
of the star formation in the far outer reaches of nearby galaxies.

\section*{Acknowledgments}

YMC is supported by the National Natural Science Foundation of China
(NSFC) under NSFC-10878010, 10633040, 11003007 and 11133001.

\label{lastpage}

\end{document}